          [2001/04/25 1.1 (PWD)]
\documentclass[a4paper,twocolumn]{esapub} 
\usepackage{natbib,graphicx}

\def\Al{$^{26}$Al}

\def\Ni{$^{56}$Ni}

\def\Na{$^{22}$Na}
\def\Co{$^{56}$Co}
\def\Ci{$^{57}$Co}

\def\Fe{$^{56}$Fe}

\def\Fh{$^{60}$Fe}

\def\Ti{$^{44}$Ti}

\def\Ca{$^{44}$Ca}
\def\Fhal{$^{60}$Fe/$^{26}$Al}

\def\ra{$\rightarrow$}
\def\aa{$\alpha$}
\def\ga{$\gamma$}

\def\cs{cm$^{-2}$ s$^{-1}$}
\def\ms{M$_{\odot}$}
\def\zs{Z$_{\odot}$}
\def\ps{e$^+$ s$^{-1}$}

\title{Astrophysical $\gamma$-ray lines: a probe of stellar nucleosynthesis
and star formation}
\author{Nikos Prantzos}
\affil{Institut d'Astrophysique de Paris, France}

\begin{document}


\maketitle

\begin{abstract}
Astrophysical gamma-ray spectroscopy is a most valuable tool
for studying nuclear astrophysics, as well as recent
star formation  in the 
Milky Way. After a short, historical, introduction to the field, 
I present a brief review of the most important current issues. 
Emphasis is given to radioactivities produced by massive stars
and associated supernova explosions, and in particular, those
related to observations presently carried out by INTEGRAL:
short-lived \Ti \ from CasA and SN1987A and long-lived
\Al \ and \Fh \ from massive stars; various candidate sources
of positrons for the 511 keV emission of the Galactic bulge
are also critically discussed.
\end{abstract}

\section{Historical overview}

The starting point of $\gamma$-ray line astronomy with cosmic radioactivities 
is usually considered to be
the landmark paper of Clayton, Colgate and Fishman (1969). That work
clarified the implications of the production of \Ni \ (a doubly magic, and 
yet unstable nucleus) during explosive Si-burning in supernovae (SN). 
In particular, it opened
exciting perspectives for $\gamma$-ray line astronomy, by suggesting that
any supernova within the local group of galaxies woud be detectable
in the characteristic $\gamma$-ray lines 
resulting from the radioactive decay of \Ni \ and its daughter nucleus \Co.

In the 70's D. Clayton identified most of the radionuclides of 
astrophysical interest (i.e. giving a detectable $\gamma$-ray line signal);
for that purpose, he evaluated their average SN yields by assuming that the
corresponding daughter stable nuclei are produced in their solar system 
abundances. Amazingly enough (or naturally enough, depending on one's point
of view) his predictions of average SN radionuclide yields (Table 2 in
Clayton 1982) are in excellent agreement with modern yield calculations, based
on full stellar models and detailed nuclear physics (see Fig. 1). Only the
importance of \Al \ escaped Clayton's (1982) attention, perhaps because its
daughter nucleus $^{26}$Mg is produced in its stable form, making the
evaluation of the parent's yield quite uncertain. That uncertainty did not
prevent Arnett (1977) and Ramaty and Lingenfelter (1977) 
from arguing (on the basis of
Arnett's (1969) explosive nucleosynthesis calculations) that, even if only
10$^{-3}$ of solar $^{26}$Mg is produced as \Al, the resulting Galactic
flux from tens of thousands of supernovae (during the $\sim$1 Myr lifetime
of \Al) would be of the order of 10$^{-4}$ \cs.

\begin{figure*}
\centering
\includegraphics[height=0.67\textwidth,angle=-90]{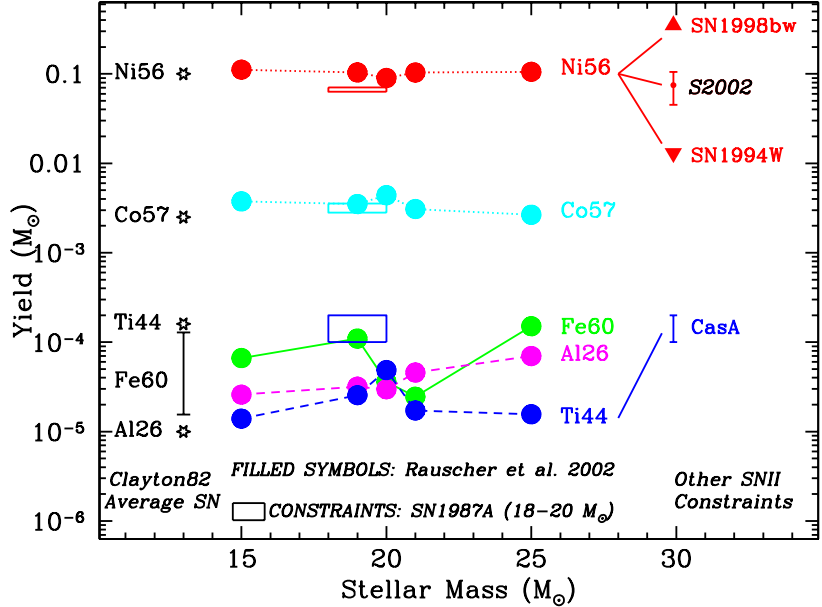}
\caption{Yields of radioactive nuclei from massive exploding stars. 
{\it On the left},  Clayton's (1982) predictions for average SN yields
are given (based on 
the assumption that the corresponding stable daughter isotopes are produced 
in their solar abundances as radioactive progenitors); the uncertainty in
\Fh \ yield stems from the unknown percentage of its contribution to the
production of stable $^{60}$Ni (which may also be produced as unstable
 $^{60}$Co; the limits of the error bar correspond to 0 percent and to
100 percent contribution, respectively). 
{\it In the middle}, recent theoretical 
results are plotted as a function of stellar mass
({\it filled symbols}, from Rauscher et al. 2002), while 
{\it open parallelograms} 
indicate observational constraints from SN1987A, a 18-20 \ms \ star.
{\it On the right} appear other observational constraints, either  from
CasA (for \Ti) or from extragalactic SN (for \Ni, see also Fig. 3).
}
\end{figure*}

In the case of \Al \  nature appeared quite generous, providing a \ga-ray
flux even larger than the optimistic estimates of Ramaty and Lingenfelter 
(1977): the HEAO-3 satellite detected the corresponding 1.8 MeV line from 
the Galactic  center direction at a level
of 4 10$^{-4}$ \cs (Mahoney et al. 1984). That detection, the first ever of
a cosmic radioactivity, showed that nucleosynthesis is still active in the 
Milky Way; however, the implied large amount of galactic \Al \ ($\sim$2 \ms \
per Myr, assuming steady state) was difficult to accomodate in conventional
models of galactic chemical evolution if SN were the main \Al \ source
(Clayton 1984), since $^{27}$Al would be overproduced in that case; however, 
if the  ``closed box model'' assumption is dropped and {\it infall}
is assumed in the chemical evolution model, that difficulty
is removed, as subsequently shown by Clayton and Leising (1987).

Another welcome mini-surprise came a few years later, when the \Co \
\ga-ray lines were detected in the supernova SN1987A, a $\sim$20 \ms \
star that exploded in the Large Magellanic Cloud. On theoretical  grounds,
it was expected that a SNIa (exploding white dwarf of $\sim$1.4 \ms \ that
produces $\sim$0.7 \ms \ of \Ni) would be the first to be detected in \ga-ray 
lines;
indeed, the large envelope mass of massive exploding 
stars  ($\sim$10 \ms) allows only small  amounts
of \ga-rays to leak out of SNII, 
making  the detectability of such objects problematic.
Despite the intrinsically weak \ga-ray line emissivity of SN1987A, the 
proximity
of LMC allowed the first detection of the tell-tale \ga-ray line signature from
the  radioactive chain  \Ni\ra\Co\ra\Fe; this confirmed a 20-year 
old
conjecture, namely that the abundant \Fe \ is produced in the
form of radioactive \Ni.

Those discoveries laid the observational foundations of the field of \ga-ray
line astronomy with radioactivities. The next steps were made in the 90ies, 
thanks
to the contributions of the Compton Gamma-Ray Observatory (CGRO). First, the
{\it OSSE} instrument aboard CGRO detected the \Ci \ \ga-ray lines from SN1987A
(Kurfess et al. 1992); the determination of the
abundance ratio of the isotopes with mass numbers
56 and 57 offered a unique probe of the physical conditions in the innermost
layers of the supernova, where those isotopes are synthesized 
(Clayton et al. 1992). On the other hand, the {\it {\it COMPTEL}} \ instrument 
mapped the
Miky Way in the light of the 1.8 MeV line and found irregular emission
along the plane of the Milky Way and prominent ``hot-spots'' in directions
approximately tangent to the spiral arms (Diehl et al. 1995); that map implies
that massive stars (SNII and/or WR) are at the origin of galactic \Al \ (as
suggested by Prantzos 1991, 1993) and not an old stellar population like
novae or AGB stars. 

Furthermore, {\it {\it COMPTEL}} \  detected the 1.16 MeV line of
 radioactive
\Ti \ in the Cas-A supernova remnant (Iyudin et al 1994). That discovery
 offered
another valuable estimate of the yield of a radioactive isotope produced
in a massive star explosion (although, in that case the progenitor star mass
is not known, contrary to the case of SN1987A). On the other hand, it also 
created some new problems, since current 1D models of core collapse supernova
do not seem able to account for the yield inferred from the observations;
however, recent axisymmetric models of rotating star explosions
offer interesting perspectives in that respect (see next section).

\begin{table*}[t]
\caption{Important stellar radioactivities for gamma-ray line astronomy
\label{fig:Table1}}
\vskip -2.8 cm
\includegraphics[width=\textwidth,height=19.cm]{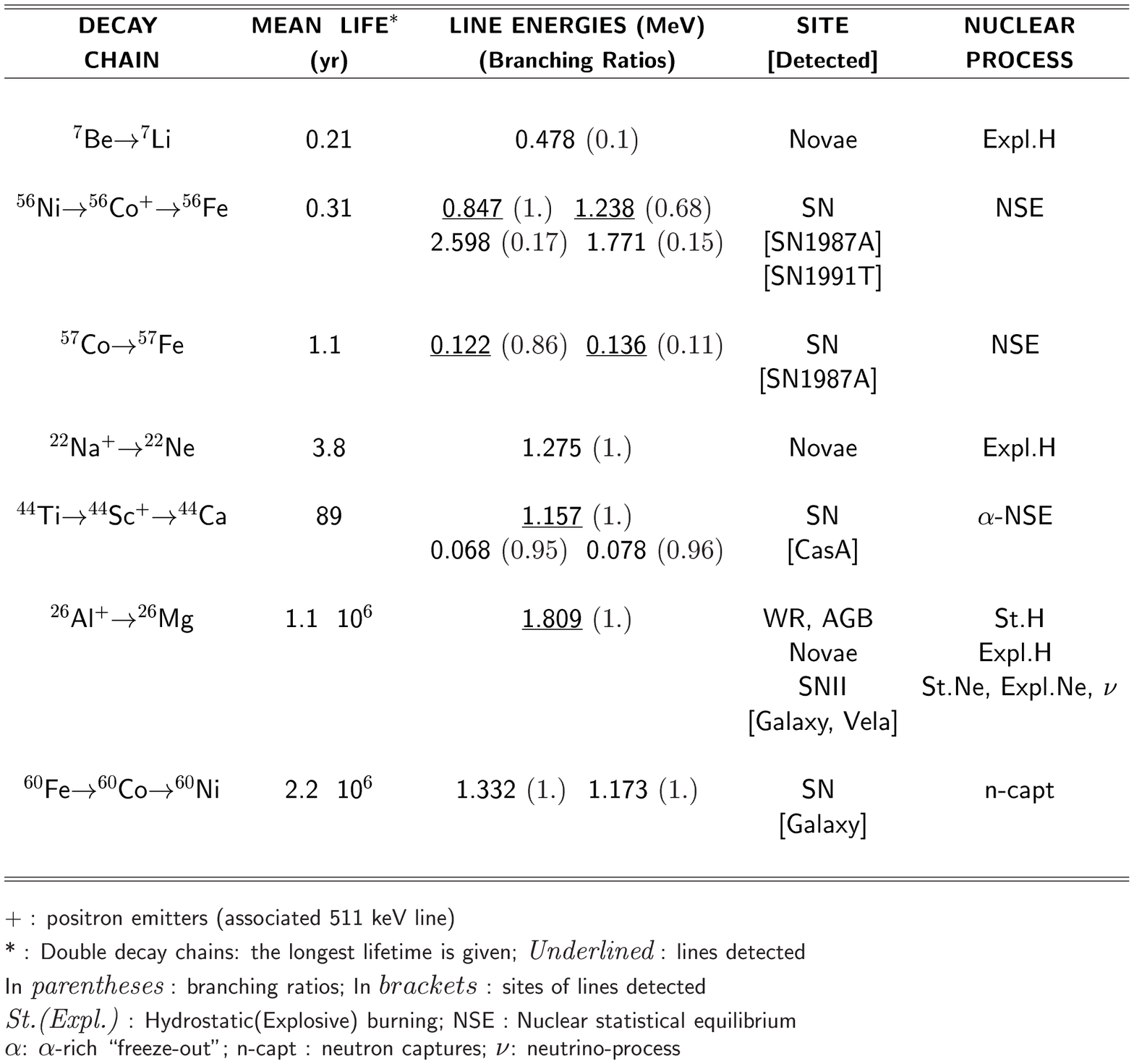}
\vskip -3.4 cm
\end{table*}

Finally, after CGRO and before INTEGRAL, another important discovery was
made in the field: the RHESSI experiment confirmed the detection of
\Al \ in the Galaxy, but it also detected the characteristic decay lines
of \Fh, another long-lived isotope with $\sim$2 Myr period
(Smith 2004, and this volume). The detected \Fh \ emissivity is, 
however,  considerably  lower than expected from current models of
nucleosynthesis in core collapse SN.  
The implications of that discovery are studied
in Prantzos (2004) and in Sec. 4.

Table 1 summarizes our current picture of stellar radioactivities and
associated $\gamma$-ray lines. In the following we shall focus on the 
radioactivities produced by massive stars and associated supernova explosions, 
and in particular, those related to observations presently carried 
out by INTEGRAL. Similar reviews have been presented by Diehl and Timmes (1998)
and Kn\"odlseder and Vedrenne (2001).
The current status of the physics of supernova explosions
is reviewed by Hillebrandt (this volume).

\section{\Ni \ and \Ti \ from core collapse supernovae}

Both \Ni \ and \Ti \ are produced in the innermost layers of core
collapse SN, through explosive Si-burning. In the high temperatures
resulting from the passage of the shock wave, matter enters Nuclear
Statistical Equilibrium (NSE). Electron captures are too slow 
to change the  neutron/proton ratio during the $\sim$1 s timescale of 
explosive nucleosynthesis and thus most of the original composition 
(consisting of $^{28}$Si, with equal numbers of protons Z and neutrons N) 
turns into \Ni \ (Z=N=28).

After the passage of the shock wave, material brought to NSE cools down. 
In environments with
relatively low densities this happens in the presence of a large concentration
of free $\alpha$ particles, which have no time to merge back to Fe-peak nuclei
through the inefficient 3-$\alpha$ reaction (the rate of which scales
with density squared). This ``$\alpha$-rich freeze-out'' process favours in
particular the production of \Ti (see discussion in 
Thielemann et al. 1996, and in particular Figs. 3 and 4a).

The yields of \Ni, \Ti \ and other Fe-peak nuclei are extremely difficult to 
evaluate from first principles, at least in the framework of current models
of core collapse supernova. The layers undergoing explosive Si-burning are 
very close to the ``mass-cut'', that fiducial surface separating the supernova
ejecta from the material that falls back to the compact object (after the
passage of the reverse shock). Since no consistent model of a core 
collapse supernova explosion exists up to now (e.g. Janka et al. 2003), 
the position of  the mass-cut
is not well constrained.  Current 1D simulations suggest that the yield of
\Ti \ is even more sensitive to the mass-cut than  the one of \Ni \ 
(see Fig. 2).

\begin{figure}
\centering
\includegraphics[height=0.49\textwidth,angle=-90]{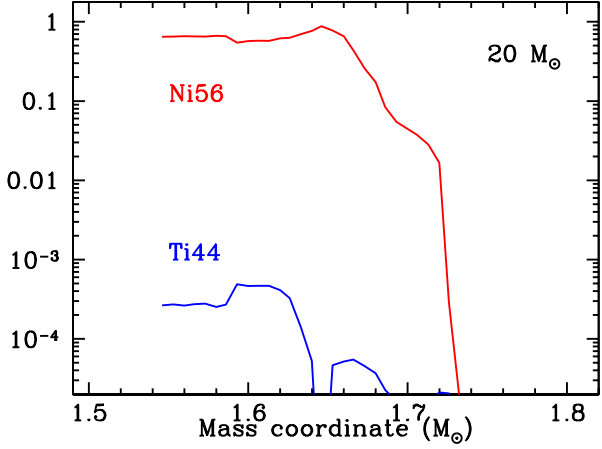}
\caption{\Ni \ and \Ti \ abundance profiles in the ejecta of a 20 \ms \ star
(from models calculated by Rauscher et al. 2002).
\label{fig:double}}
\end{figure}

The presence of \Ni \ in SN1987A has been unambiguously
inferred from the detection of
the $\gamma$-ray lines of the decay of its daughter nucleus \Co.
The yield of \Ni \ has been estimated from the extrapolation of the early
optical lightcurve to the origin of the explosion (precisely known thanks
to the neutrino signal, see Arnett et al. 1989 and references therein);
the derived value, 0.07 \ms, is often taken as a ``canonical'' one for
core collapse SN.

It turns out, however, that core collapse SN display a wide range of \Ni \ 
values, spanning a range  of at least one order of magnitude, as can be seen in
Fig. 3 (Hamuy 2003 and references therein). 
Despite the large error bars, it seems that there is a clear 
correlation between the amount of \Ni \ and the energy of the explosion
(obviously, because a shock of larger energy heats a larger amount of
material to NSE conditions).

\begin{figure}[b!]
\centering
\includegraphics[width=0.49\textwidth]{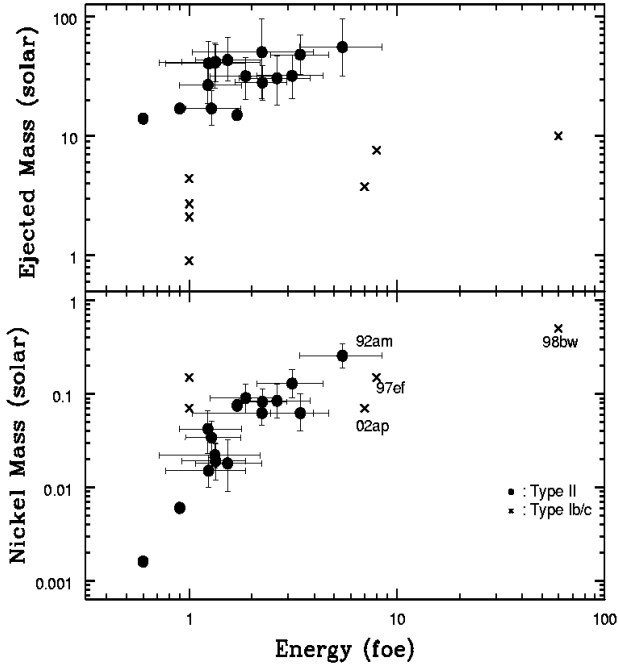}
\caption{Total ejected mass (upper panel) and \Ni \ yield (lower panel)
as a function of ejecta kinetic energy for core collapse supernovae
(filled dots: SNII, crosses SNIb/c), according to Hamuy (2003).
\label{fig:double}}
\end{figure}

\begin{figure}
\centering
\includegraphics[height=0.48\textwidth,width=0.35\textwidth,angle=-90]
{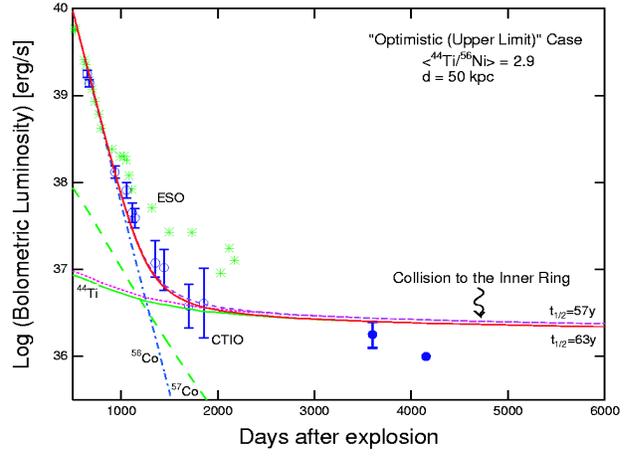}
\caption{Contributions of various radioactivities to the
observed UVOIR lightcurve of SN1987A (from Motizuki and Kumagai 2003a).
The late lightcurve requires the presence of $\sim$2 10$^{-4}$ \ms \ of
\Ti; the derived \Ti/\Ni \ ratio is $\sim$ 3 times higher than the
solar ratio of the corresponding stable isotopes 
($^{44}$Ca/$^{56}$Fe)$_{\odot}$.
\label{fig:double}}
\end{figure}

\Ti \ has not been directly detected in SN1987A up to now. Modelling of the
late lightcurve of that supernova suggests that it may be powered by 
1-2 10$^{-4}$ \ms \ of \Ti \ (Motizuki and Kumagai 2003a) 
and similar values are
obtained through analysis  of the infrared emission line of the ejecta
(Fransson and Kozma 2002). Note that the evaluation of the \Ti \ yield 
through these methods  suffers from considerable uncertainties, due to the
complex physics of the supernova heating and cooling and the role of positrons.
The derived amount (and especially the higher value)
is rather high compared to the
results of 1D models of 18-20 \ms \ stars (e.g. from Rauscher et al. 2002,
see Fig. 1). Moreover, the \Ti/\Ni \ mass ratio is around 3 times
the solar ratio of the corresponding stable isotopes
($^{44}$Ca/$^{56}$Fe)$_{\odot}$, too
high to be explained by current 1D models, as will be discussed below.

Despite these high values, \Ti \ from SN1987A seems beyond the detection
capabilities of INTEGRAL, now that the on-flight performance of {\it SPI} is
well established. The analysis of Motizuki and Kumagai (2003a) indicates that
the expected
flux in the 1157 keV line is $\sim$5 10$^{-6}$ ph/cm$^2$/s, i.e. considerably
lower than the $\sim$2 10$^{-5}$ ph/cm$^2$/s sensitivity of {\it SPI} for an
exposure of 1 Ms.

The $\gamma$-ray lines of \Ti \ have been detected in the $\sim$320 yr old
CasA supernova remnant, lying 
at a distance of $\sim$3.4 kpc from earth. Both the
high energy line at 1.157 MeV and the low energy ones, at 68 and 78 keV,
have been detected, respectively by {\it COMPTEL} (Iyudin et al. 1994) and
Beppo-SAX (Vink et al. 2001). 
The detected flux of 3.3$\pm$0.6 10$^{-5}$ ph/cm$^2$/s
from {\it COMPTEL},  points to a \Ti \ yield
of $\sim$1.7 10$^{-4}$ \ms. Similar values, i.e. 1-2 10$^{-4}$ \ms, are 
obtained through a study of the combined fluxes of the low energy lines
(Vink and Laming 2002), although the modelisation of the underlying continuum
spectrum makes the analysis very difficult.

\begin{figure}[t!]
\centering
\includegraphics[height=0.48\textwidth,angle=-90]{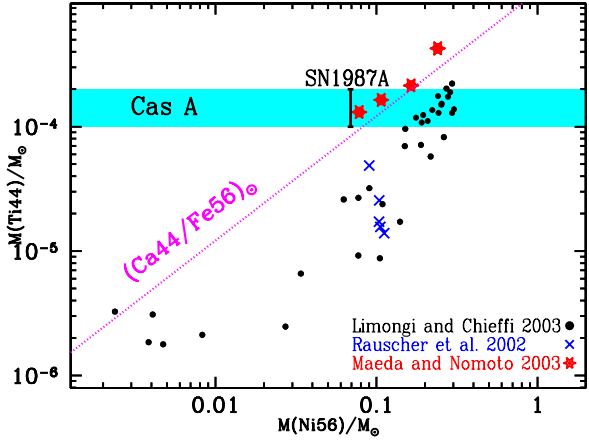}
\caption{Yield of \Ti \ vs yield of \Ni, from models and observations.
Model results are from Limongi and Chieffi (2003, filled dots, with 
large variations in yields due to variations in both stellar mass - from
15 to 35 \ms \ - and
explosion energy), Rauscher et al. (2002, crosses, for stars in the
15 to 25 \ms \ range and explosion energies of 10$^{51}$ ergs) and
Maeda and Nomoto (2003, asterisks, for axisymmetric explosions in
25 and 40 \ms \ stars, producing high \Ti/\Ni \ ratios). 
\Ti \ detected in CasA appears as
a horizontal shaded band (assuming that its decay rate has not been
affected by ionisation in the CasA remnant, otherwise its abundance should be
lower, according to Motizuki et al. 1999). 
The amount of \Ti \ in SN1987A is derived
from its late optical lightcurve (Motizuki and Kumagai 2003, see Fig. 4).
The diagonal dotted line indicates the solar ratio of the corresponding
stable isotopes ($^{44}$Ca/ $^{56}$Fe)$_{\odot}$.  
\label{fig:double}}
\end{figure}

The \Ti \ yield of CasA is comparable to the one implied by the
optical and infrared observations of SN1987A. Note, however, that the CasA
yield suffers from uncertainties related to the ionisation stage of the
CasA remnant. \Ti \ decays by orbital electron capture and an ionised
medium could slow down its decay (Mochizuki et al. 1999); the derived yield
could be considerably smaller in that case (Motizuki and Kumagai 2003b).
On the other hand, ionisation is expected to play a small role in the 
aformentioned derivation of the \Ti \ yield in SN1987A, but the remnant of 
that SN is rapidly evolving (Michael et al. 2002); if \Ti \ enters a high 
ionisation  stage in the near future, the expected $\gamma$-ray line fluxes
from SN1987A
should be accordingly reduced (see Motizuki and Kumagai 2003b).

In summary: from optical observations we have a wide range of values for the
\Ni \ yields of core collapse SN, and a precise value of 0.07 \ms \ 
for SN1987A; and for \Ti \ yields we have similar values, i.e. 1-2 10$^{-4}$
\ms,  for both SN1987A (indirectly, through the modelisation of the UVOIR 
light)  and for CasA (directly, through $\gamma$ ray lines,
albeit with a systematic uncertainty resulting from poorly constrained
ionisation effects).
How do these results compare with theoretical expectations?

The most detailed relevant calculations have been recently 
performed for 1D models, concerning stars in the 15-35 \ms \ range and
parametrised by the energy of the explosion (Limongi and Chieffi 2003, LC03).
Also, Rauscher et al. (2002, RHHW02) 
investigated a more limited range of stellar
masses (from 15 to 25 \ms), but they considered only a fixed ``canonical'' 
value for the kinetic energy of the explosion; their results supersede those
of Woosley and Weaver (1995), obtained with the same stellar evolution code
but with an older set of nuclear reaction rates. 
Both calculations of RHHW02 and
LC03 are made with the same 
reaction rate libraries and concern stars of solar initial metallicity. It
is questionable to what extent they apply to SN1987A, the progenitor of
which presumably had an LMC metallicity of 0.3 \zs.

The results of those calculations are plotted as \Ti \ yield vs \Ni \ yield
in Fig. 5, where the solar
ratio of the corresponding stable isotopes is also displayed as a diagonal 
line. It can be seen that:

\begin{figure}[t!]
\centering
\includegraphics[height=0.48\textwidth,width=0.35\textwidth,angle=-90]
{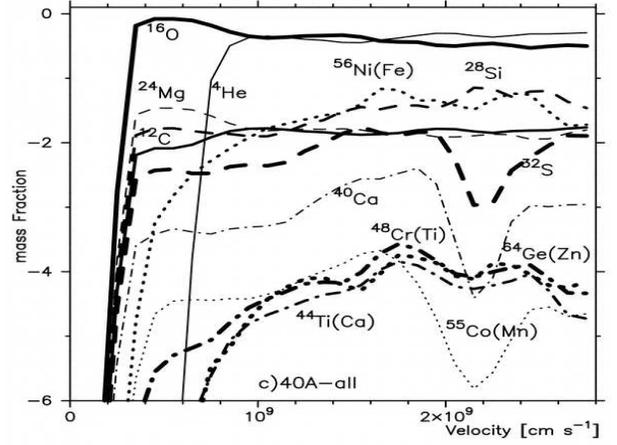}
\caption{Abundance profiles of major isotopes as a function of ejecta velocity,
after the bipolar
explosion of a rotating star of 40 \ms; profiles have been averaged
over all angles 
(from Maeda and Nomoto 2003).
\label{fig:double}}
\end{figure}

- There is broad agreement between LC03 and RHHW02, at least around the
value of 0.07-0.1 \ms \ of the \Ni \ yield.

- Except for the lowest \Ni \ yields, \Ti \ and \Ni \ yields of LC03 
both increase with the energy of the explosion, at a quasi-constant average 
ratio of $\sim$3 10$^{-4}$.

- This value is $\sim$3 times lower than the solar ratio of 
(\Ca/\Fe)$_{\odot} \sim$10$^{-3}$. This 
implies that such explosions cannot produce the solar \Ca, since
\Fe \ would be overproduced in that case (e.g. Timmes et al.
1996). Moreover, there is another important source of Fe,
SNIa, which  produce about 0.5-0.65 of solar
\Fe, but very little \Ca; this makes the defficiency of \Ca \ from core 
collapse SN even more serious than appearing in Fig. 5, since it implies
that core collapse  explosions {\it should} 
produce a \Ti/\Ni \ ratio at least {\it twice} 
solar in order to compensate for the \Fe \ production of SNIa. Such a high
ratio is also what is required to explain the high value of the \Ti \ yield
in SN1987A.

- In the case of CasA, one has two choices. The first is to assume a very 
energetic  explosion and, according to 1D models,  
get a \Ti \ yield compatible with CasA observations;
but then a high \Ni \ yield of $>$0.2 \ms \ is implied by those same models, 
leading to the question ``why such a bright explosion as CasA went undetected
back in the 1680s?'' (e.g. Hartmann et al. 1996). 
The second option is to assume 
that the \Ni \ yield was much smaller than suggested by 1D models
producing high \Ti \ (thus explaining why CasA went undetected);
in that case it becomes clear that current 1D models are  
unable to explain the situation in either CasA or SN187A.

A solution to this problem of ``missing \Ti \ (and $^{44}$Ca)'' is 
provided by multi-dimensional models of the energetic explosions
of rotating massive stars (hypernovae). 
In such explosions, material falling onto the
central remnant forms an accretion disk, and  a fraction of that material is 
ejected in the form of two jets (or a bipolar wind), which interact with
the stellar mantle. According to Nagataki et al. (1998), material along the
jet axis undergoes higher temperatures and entropies (i.e. lower densities)
than material in normal spherical explosions. Such conditions are particularly
favourable to an ``$\alpha$-rich freeze-out'' and lead to the production
of large \Ti \ amounts and  \Ti/\Ni \ ratios. In Fig. 5 it can be seen that
recent nucleosynthesis results obtained in the framework of such
models (from Maeda and Nomoto 2003) may cure all the aforementioned problems.

At least in the case of SN1987A, there is convincing evidence that the
explosion was axisymmetric and not spherical (see Wang et al. 2002 and 
references therein). On the other hand, \Ti \ produced in the jets is ejected
at much higher velocities than in spherical symmetric explosions, reaching
up to 25000 km/s according to Maeda and Nomoto (2003 and Fig. 6). This makes
its lines rather broad and their detection by a spectrometer like {\it SPI}
even more difficult. For that reason, detection of the \Ti \ lines 
from CasA and measurement of their width is
one of the most critical issues in $\gamma$-ray line astronomy today,
and one of the prime objectives of {\it SPI}/INTEGRAL.

Finally, continuing search for \Ti \ lines in the Milky Way could
reveal young supernova remnants and put interesting statistical
constraints on \Ti \ yields and supernova frequencies. Such a search has
been already performed in the past with HEAO-3, SMM and {\it COMPTEL} data;
a Monte Carlo analysis by The et al. (2000) showed that a rather rare
supernova type with high \Ti \ yield is favoured by the absence of
a \Ti \ signal from the inner Galaxy. Similar conclusions are reached
with preliminary INTEGRAL data (see contribution by Renaud et al. 
this volume).

\section{\Al \ and \Fh \ from massive stars}

\Al \ is the first radioactive nucleus ever detected in the Galaxy
through its characteristic gamma-ray line signature, at 1.8 MeV
(Mahoney et al. 1984). Taking into account that its lifetime of $\sim$1
Myr is short w.r.t. galactic evolution timescales, 
its detection convincingly demonstrates that nucleosynthesis
is still active in the Milky Way (Clayton 1984). The detected flux
is $\sim$4 10$^{-4}$ cm$^{-2}$ s$^{-1}$ and  corresponds to $\sim$2 \ms \
of \Al \ currently present in the ISM (and produced per Myr,
assuming a steady state situation). 

The {\it COMPTEL} instrument aboard
CGRO mapped the 1.8 MeV emission in the Milky Way and found it to
be irregular, with prominent "hot-spots" probably associated
with  the spiral arms (Diehl et al. 1995). The spatial distribution of \Al \
suggests that massive stars are at its origin (Prantzos 1991,
Prantzos and Diehl 1996), since any population of long-lived objects
(AGB stars or novae) is expected to be more smoothly distributed.

 \begin{figure}
 \centering
 \includegraphics[height=0.49\textwidth,angle=-90]{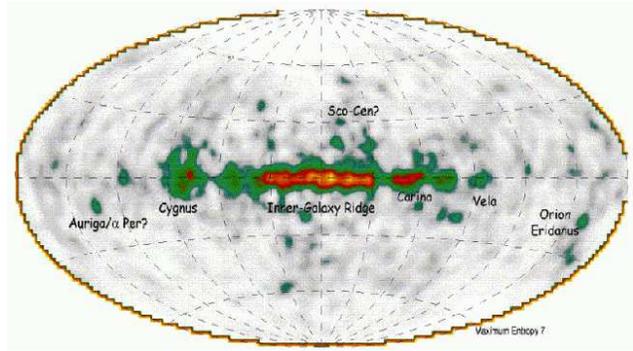}
 \caption{Map of Galactic \Al, after 9 years of {\it COMPTEL} operation 
(Pl\"uschke et al. 2001a).
\label{fig:double}}
 \end{figure}

The detailed mapping of the Galactic distribution of \Al, obtainable through
a determination of the distances to the ``hot-spots'' 
(see Kretschmer et al. 2003),  is one of the main
long-term objectives of INTEGRAL, since it will provide the most
accurate picture of recent star formation in the Milky Way (see contributions
by Diehl et al. and Hartmann et al., this volume). 
On the other hand, the study of individual 
``hot-spots'' revealed by {\it {\it COMPTEL}} \ bears 
on our understanding of the evolution of young stellar associations 
(in the cases of Cygnus, Carina 
and  Centaurus-Circinus) and even individual stars (in the case of Vela).

The Cygnus region was studied with population synthesis models by two groups 
(Cervinho et al. 2001,  Pl\"uschke et al. 2001b). The
resulting morphology of the 1.8 MeV emission compares well with the 
{\it {\it COMPTEL}} \ data.
However, in the case of Carina, the predicted absolute flux 
is smaller (by a factor of 5-20) than detected by
{\it {\it COMPTEL}} (Kn\"odlseder et al. 2001). 
That discrepancy may imply something interesting, either for the
(in)completeness of the stellar census of that association or for the
\Al \ yields. {\it INTEGRAL} will  establish more accurately the morphology of 
those ``hot-spots'' and further test the ``massive star group'' origin of \Al
(see contribution by Kn\"odlseder et al.  in this volume).

The Vela region offers the  opportunity to measure (or put upper  limits
on) \Al \ yields from individual sources. The morphology of the rather extended
1.8 MeV emission detected by {\it COMPTEL} (Diehl 2002)
does not allow identification  with any of the
three known objects in the field (the Vela SNR, the closest WR star \ga$^2$ 
Vel and 
SNR RX-J0852-4622); all three objects lie closer than 260 pc, according
to recent estimates. {\it {\it COMPTEL}} \ measurements are  compatible with 
current yields of SNII (in the
case of Vela SNR) and marginally compatible with current yields of \ga$^2$ Vel 
(Oberlack et al. 2000). INTEGRAL measurements in the Vela region are then
expected to place more stringent constraints on stellar 
nucleosynthesis models (see 
contribution by Mowlavi et al. in this volume).

\begin{figure}
\centering
\includegraphics[height=0.49\textwidth,angle=-90]{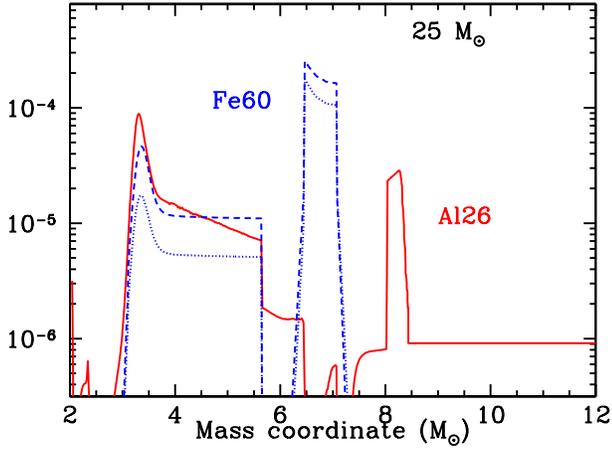}
\caption{Radial profiles of \Al \ (solid curve) and \Fh \ (dashed curve)
in a 25 \ms \ star (from Rausher et al. 2002). The dotted curve is also for
\Fh \ and it is obtained by reducing by a factor of 2 the cross section of
$^{59}$Fe(n,$\gamma$)\Fh \ (Al. Heger, private communication).  
\label{fig:double}}
\end{figure}

It is not yet clear
whether the majority of observed \Al \ originates from the winds
of the most massive stars (i.e. above 30 \ms, evolving as
Wolf-Rayet stars) or from the explosions of less massive stars
(i.e. in the 10-30 \ms \ range, exploding as SNII); the
uncertainties in the corresponding stellar yields are still quite
large and do not allow to conclude yet.

Clayton (1982)  pointed out that SNII explosions produce
another relatively short lived radioactivity, \Fh \ (lifetime
$\sim$2 Myr). Since WR winds do not eject that isotope, the
detection of its characteristic gamma-ray lines   in the
Milky Way would constitute a strong argument  for SNII being at
the origin of \Al. Based on detailed nucleosynthesis calculations
of SNII (from Woosley and Weaver 1995, hereafter WW95) 
Timmes et al. (1995) found
that the expected gamma-ray line flux ratio of \Fhal \
(for each of the  two lines of \Fh) is
0.16, if SNII are the only sources of \Al \ in the Milky Way.

The Reuven Ramaty High Energy Solar Spectroscopic Imager (RHESSI)
detected  the galactic \Al \ emission at a flux level compatible
with previous observations (Smith 2003). Recently, Smith
(2004) reported the first ever detection of the Galactic \Fh \
gamma-ray lines with RHESSI.
A reanalysis of the data (see contribution by 
Smith, this volume) leads to a marginal significance of the
detection (2.6 $\sigma$) and to a  line flux ratio \Fhal \ of 0.10 (for each
\Fh \ line), i.e. slightly lower than  predicted by Timmes et al
(1995) on the basis of  WW95 nucleosynthesis calculations.

However, more recent studies of SNII nucleosynthesis (Rauscher et al. 2002,
hereafter RHHW02; and Limongi and Chieffi 2003, hereafter LC03)
produce different yields than WW95 (more \Fh \ and less \Al); in particular,
a large amount of \Fh \ is produced in the He-layer (Fig. 8),
probably due to the different nuclear reaction rates used in the new
calculations. As a result,  the \Fhal \ ratio turns out to be 
considerably higher than the one found in Timmes et al. (1995). 

\begin{figure}[t!]
\centering
\includegraphics[height=0.49\textwidth,angle=-90]{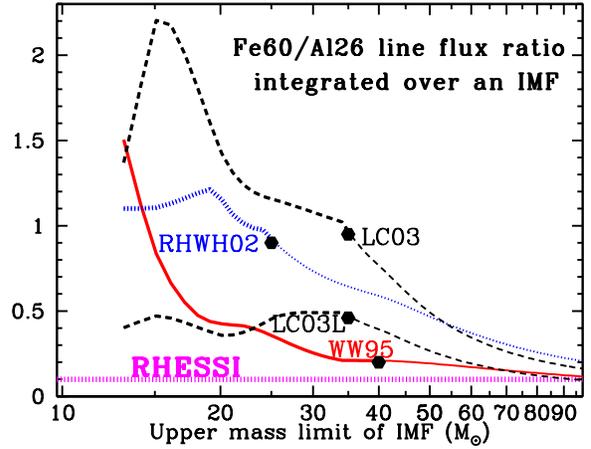}
\caption{The expected ratio of \Fh/\Al \ decays
(for each of the two \Fh \ lines), convolved with a Salpeter
stellar Initial Mass Function, is shown as a function of the upper
stellar mass limit of the convolution integral. The
four curves correspond to the four different sets of stellar
yields, with their thick portions corresponding to the mass range
covered in those works (see text). The {\it dotted } horizontal
line at 0.10 is the \Fh/\Al \ line flux ratio reported by
RHESSI (Smith, this volume) or, perhaps, an upper limit on that ratio. 
{\it Filled pentagons} mark the upper mass in each of the
four studies; all recent calculations predict much higher values
than the older calculations of WW95 or the value of
RHESSI. Only by taking into account the \Al \ yields of massive
Wolf-Rayet stars ({\it thin} portion of the curves beyond the masses
indicated by the filled pentagons,  obtained
by adding data on \Al from WR stars from Meynet et al.
1997) one may obtain \Fh/\Al \ ratios
compatible with the RHESSI results.
\label{fig:double}}
\end{figure}

Prantzos (2004) compared the new yields  with observations, after
convolving with a stellar Initial Mass Function (IMF). The
results are plotted in Fig. 9. In the case of LC03 two curves are
shown: LC03H corresponds to the high \Fh \ yields (low explosion
energies) and LC03L to the low \Fh \ yields (high explosion
energies). In all cases the thick portions of the curves
correspond to the stellar mass range covered by each study, while
the filled hexagons mark the highest mass of each calculation and
thus provide the IMF weighted value over the whole mass range
covered by each study.
It is clear from Fig. 9 that recent calculations are in disagreement with
the RHESSI result.

\begin{figure}
\centering
\includegraphics[height=0.49\textwidth,angle=-90]{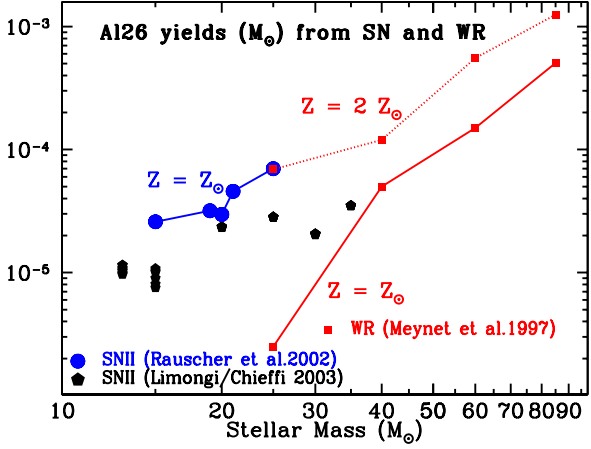}
\caption{\Al \ yields of massive stars, either ejected by SNII explosions
(from Rauscher et al. 2002 and Limongi and Chieffi 2003) or expelled from
WR winds (from Meynet et al. 1997). SNII yields are for stars of 
metallicity \zs, while those of WR  for metallicities \zs and 2 \zs; the
latter is more appropriate for the average metallicity of the Milky Way.  
\label{fig:double}}
\end{figure}

Assuming that the recent theoretical results are not to be
substantially revised in the future and that the RHESSI result is
confirmed,what are the implications for our understanding of the
origin of \Al? The obvious conclusion is that the bulk of galactic
\Al, detected by various instruments including RHESSI, is not
produced by the source of \Fh: if this were the case, then \Fh \
would be detected with a line flux similar to the
one of \Al. Obviously, another source
of \Al \ is required, producing much smaller \Fhal \ ratios
than the SNII.

The obvious candidate source is Wolf-Rayet stars, as has been
argued in many places over the years (e.g. Dearborn and Blake
1985, Prantzos and Cass\'e 1986, Prantzos 1993, Prantzos
and Diehl 1996, Meynet et al. 1997, Kn\"odlseder 1999). The winds
of those massive, mass losing stars, eject large amounts of \Al \
produced through H-burning in the former convective core, {\it
before} its radioactive decay. In stars with no mass loss, those
quantities of \Al \ decay inside the stellar core before the final
explosion and never get out of the star. Unfortunately, no consistent
model for the complete evolution of a massive, mass losing,  
star (i.e. up to the final explosion) exists up to now.

Adopting the \Al \ yields of non-rotating WR stars of solar
initial metallicity by Meynet et al (1997), which concern stars 
in the 25-120 \ms \ mass range, and combining
them with the aforementioned SNII yields (see Fig. 10), 
one obtains the \Fh/\Al \ line flux
ratio produced by the total mass range of massive stars, during
all the stages of their evolution; this is expressed in Fig. 9 by
the continuation of the four theoretical curves above the masses
indicated by the filled pentagons. It can be seen that the RHESSI
result is recovered in that case, provided that at least half of
\Al \ originates from WR stars (in the case of LC03L), or even
that 80\% of \Al \ originates from WR stars (in the case of RHHW02
or LC03H).

\begin{figure}[t!]
\centering
\includegraphics[height=0.49\textwidth,angle=-90]{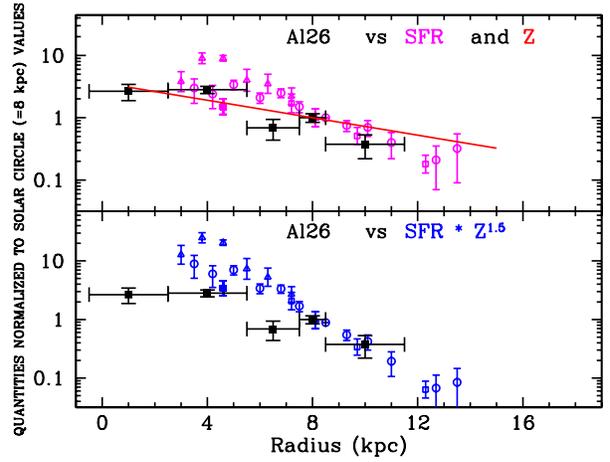}
\caption{Radial distributions of Al26, star formation rate
(SFR) and metallicity (Z) in the Milky Way disk. {\it Upper panel:}
Data points with vertical error bars correspond to various tracers of the
SFR, while the galactic metallicity profile of oxygen (with
a gradient of dlog(O/H)=-0.07 dex/kpc) is shown by a solid line;  the
Al26 profile, after an analysis of {\it {\it COMPTEL}} \ data by Kn\"odlseder
(1997), is shown  (in relative units)
by data points with vertical and horizontal error bars (the horizontal ones
correspond to the adopted radial binning). {\it Lower panel:} If galactic
Al26 originates mostly from WR stars, its radial distribution should scale
with SFR * Z$^{1.5}$ (points with vertical error bars, scaled from the upper
panel), since the Al26 yields of WR stars scale with Z$^{1.5}$ at least
(Vuissoz et al. 2003, calculations for rotating stars);
however, the observed Al26 distribution
(same points as in upper panel) is flatter than the expected
one in that case.
\label{fig:double}}
\end{figure}

Those arguments point towards WR stars as major sources of \Al \
in the Milky Way. However, the situation  is far from  clear
yet, because the WR stellar yields of \Al \ depend strongly on
metallicity. In the case of non rotating stellar models that
dependence is $\propto$ Z$^2$, according to Meynet et al. (1997).
The rotating models of WR stars, currently calculated by the
Geneva group (Meynet and Maeder 2003) show that rotation
considerably alleviates the need for high mass loss rates, while
at the same time leading to the production of even larger \Al \
yields than the non-rotating models (Vuissoz et al. 2003, Palacios et al.
this volume); in that
case, it is found that the \Al \ yields of WR have a milder
dependence on metallicity ($\propto$ Z$^{1.5}$) than the non
rotating ones. 

In both cases, that metallicity dependence of the
\Al \ yields of WR stars, combined with the radial profiles of star
formation rate (SFR) and of metallicity in the Milky Way (see Fig.
11, upper panel) suggest that the resulting radial profile of \Al \
should be much steeper than the one actually observed. The latter,
derived from {\it COMPTEL} observations (Kn\"odlseder 1997) appears in
Fig. 11 (lower panel) and is clearly flatter than the product
SFR*Z$^{1.5}$ (as already noticed in Prantzos 2002). Similar
conclusions are reached if the longitude, rather than radial,
profiles of \Al, metallicity and SFR  are considered.

Thus, almost twenty years after its discovery,
the \Al \ emission of the Milky Way has not yet found a
completely satisfactory explanation. Indeed, the recent
observational ({\it COMPTEL}, RHESSI) and theoretical (RHHW02, LC03,
Vuissoz et al. 2003) results have made the puzzle even more
complex than before. The solution will obviously require progress
in both theory and observations. From the theory point of view, detailed
nucleosynthesis calculations of mass losing and rotating stars up
to the final explosion in the mass range 12-100 \ms \ and for
metallicities up to 3 \zs \ will be required ; furthermore,
the uncertainties still affecting the reaction rates of $^{22}$Ne($\alpha$,n) 
(major neutron producer
during He burning in massive stars) and $^{59}$Fe(n,$\gamma$)\Fh \ 
will have to be substantially reduced. From the observational
point of view, the radial distributions of both \Al \ and \Fh  \
will be needed; such distributions will probably be available if
the operation of INTEGRAL  is prolonged for a few
years beyond its nominal 2-year mission (see contribution by C. Winkler,
this volume).

\section{Positron annihilation in the Galaxy}

The first $\gamma$-ray line ever detected outside the solar system was the
511 keV line of electron-positron annihilation (Johnson et al. 1972, Leventhal
et al. 1978). Observations by various instruments in the 90's
established that the line is not variable (at least in a $\sim$10 year period),
that its spatial distribution is apparently dominated by a bulge-like
component and that the overall spectrum suggests a large positronium
fraction of 0.93 (see Kinzer et al. 2001 and references therein). 
The 511 keV flux detected by $\gamma$-ray spectrometers 
(HEAO-3, GRIS, HEXAGONE, TGRS) in the central Galactic sterad
was found to be $\sim$10$^{-3}$ ph/cm$^2$/s, corresponding
to a steady state production rate of 10$^{43}$ \ps.

Observations during the early mission phase of INTEGRAL broadly confirm that
picture. Analysis of {\it SPI} data (Kn\"odlseder et al. 2003, 
Jean et al. 2003,
contribution by Jean et al. this volume)
finds a rather narrow line (FWHM$\sim$3 keV),
excludes a point source in the Galactic Center (GC),
and finds that the emission is bulge-like (best fit with a 
Gaussian of FWHM=9$^o$).
The non-detection of a disk emission up to now imposes a model-dependent
lower limit on the bulge/disk ratio of 0.4-0.8 
(see also contribution by Weidenp\"ointer et al. this volume).
The detected flux of  9.9$^{+4.7}_{-2.1}$ 10$^{-4}$ ph/cm$^2$/s 
is compatible with  previous measurements.

Possible sources of Galactic positrons have been studied in several 
works up to now
(e.g. Dermer and Murphy 2001 and references therein). 
Most prominent among them seems to be the $\beta^+$-radioactivity 
of supernovae (see Table 1) and in particular of SNIa. On the other hand, 
the already detected emission of \Al \ in the Galaxy (Sec. 3) implies that
this radioactive nucleus contributes a non-negligible fraction of the 
Galactic positron production rate. 
A thorough analysis of the supernova contribution to positron production 
in the Galaxy 
has been done by Milne et al. (1999a). Their conclusion was that the decay 
of \Co \ from
the {\it total} Galactic population of SNIa may produce $\sim$ 10$^{43}$ \ps,
and that this rate could explain about half of the
the total Galactic  511 keV  emission as measured
by OSSE, SMM and TGRS.

The improved spatial (and spectral) resolution of {\it SPI} gives 
a dramatic new twist
to the issue of Galactic positrons, because it is clear now that 
$\sim$10$^{43}$ \ps \
are required to be produced in the bulge {\it alone}, not in the whole
Milky Way (unless a mechanism is found 
to channel all positrons from the disk to the bulge).

The positron production rate from SNIa in the bulge is
$$ R \ = \ M \ F \ N $$
where: $M$ is the mass of the bulge, $F$ is the SNIa frequency per unit mass in
a bulge-like system, $N$ is the number of positrons ejected by a typical
SNIa. Typical values for those parameters are: 
$M$=1.5$^{+0.5}_{-0.5}$ 10$^{10}$
\ms \ (e.g. Launhardt et al. 2002), $F$=0.22$^{+0.07}_{-0.07}$ SNIa per 
10$^{10}$ \ms \ per millenium (Cappellaro et al. 2003, assuming that 
the bulge is of intermediate type, between E0 and Sa) and $N$=8$^{+7}_{-4}$
10$^{52}$ positrons (Milne et al. 1999b, i.e. 
the \Co \ positron escape fraction
is $\sim$3\%). With those numbers one gets $R \sim$ 0.8 10$^{42}$ 
\ps, i.e. a production rate about 
12 times lower than suggested by the {\it SPI} data analysis.
The expected contribution from other radioactivities (\Ti \ from SN, \Al \ from
massive stars, \Na \ from novae) is too low to account for the missing amount
of positrons.

\begin{figure}[t!]
 \centering
 \includegraphics[height=0.49\textwidth,width=0.39\textwidth,angle=-90]{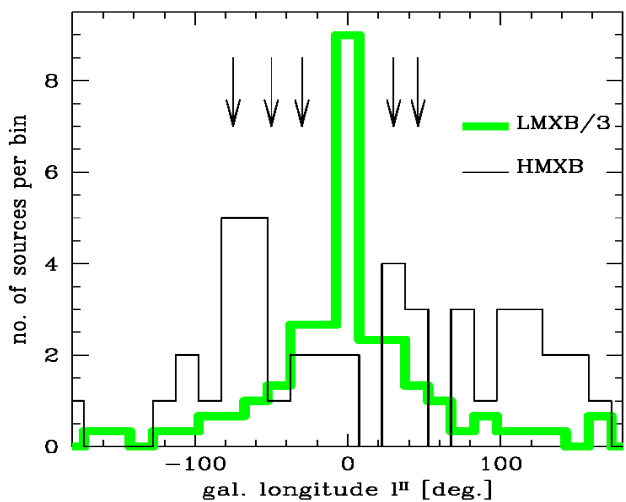}
 \caption{Longitude distribution of galactic LMXBs (thick histogram) and HMXBs (thin
histrogram), from Grimm et al. (2002). The collective emissivity of LMXBs is
$\sim$10 times higher than the one of HMXBs and 100 times higher than required
to produce 10$^{43}$ \ps.
\label{fig:double}}
 \end{figure}

It should be stressed that the uncertainties on $F$ and $N$ could be 
considerably
larger than quoted in the previous paragraph. Indeed, the fact that SNIa have not been
detected up to now in the bulges of spirals (Wang et al. 1997) suggests that 
important selection biases, e.g. extinction, may affect the results of current surveys.

\begin{table*}
\caption{Candidate sources for positrons in Galactic bulge
\label{fig:Table2}}
\begin {center}
\begin{tabular}{cccc}
\hline \hline
{\bf SOURCE} & {\bf MORPHOLOGY} & {\bf INTENSITY} & {\bf COMMENTS} \\
\hline
     &       &    &          \\
{\bf SNIa}   & Consistent with & $\sim$10 times low                    &  Uncertainties        \\
{\bf radioactivity}& observations    & (``standard'' parameter values)   &  (perhaps) underestimated  \\
     &       &    &          \\
 {\bf LMXB}             & Seems OK, BUT       & Unknown, but   &  Less than 1\% of available       \\
 {\bf Outflows/jets} & dominated by few strong      & does not seem unreasonable   & energy required   \\
    & sources outside the bulge      & because $\longrightarrow$   &  to form positrons       \\
     &       &    &        \\
    & Improbable       & Not unreasonable,   & More hypernovae         \\ 
 {\bf Hypernova(e) }    & (how to fill the bulge      & but very poorly & expected in disk     \\
     & with positrons?)      & constrained  & (molecular ring)         \\
     &       &    &        \\
{\bf Dark Matter}     & Centered on GC,       & Unknown/   &    Postulated light particle    \\
{\bf annihilation}     & otherwise unknown       & Unconstrained   & should be seen in LEP     \\
     &       &    & (if in Standard Model)       \\
     &       &    &        \\
\hline \hline

\end{tabular}
\end{center}
\end{table*}

On the other hand, the value of $N$ has been empirically derived from the late 
lightcurves of SNIa (Milne et al. 1999b). This estimate is on much more secure grounds
than the theoretical estimates of Chan and Lingelfelter (1993). However, it may
also be affected by uncertainties on any early escape of positrons from
\Ni \ mixed in the outer layers. Thus, SNIa should not yet be definitively
excluded as major sources of bulge positrons. However, other positron
sources should be sought.

Among the various candidate sources, Low Mass X-Ray Binaries (LMXBs)
appear quite promising, at least as far as the morphology and the intensity constraints
of the observed 511 keV emission are concerned. Indeed, the observed distribution
of LMXBs in the Milky Way is strongly concentrated towards the Galactic bulge
(Fig. 12),
and satisfies the bulge/disk ratio obtained by the preliminary $SPI$ analysis (P. Jean,
private communication). Moreover, the collective X-ray emissivity of  Galactic 
LMXBs is $\sim$10$^{39}$ erg/s (Grimm et al. 2002), 
compared to $\sim$10$^{37}$ erg/s required to
produce $\sim$10$^{43}$ \ps \ with energies of $\sim$1 MeV per positron, 
as those
produced from radioactivity. It is sufficient to convert only $\sim$1 \% of the
available energy into positrons to explain the bulge emission. However, the
mechanism of that conversion is not known yet. Moreover, the observed X-ray
emissivity is dominated by a dozen strong sources, which are mostly
distributed in the Galactic plane and not in the bulge (see Table 5 in Grimm
et al. 2002).

Even more speculative is the suggestion (Nomoto et al. 2001) 
that very energetic
(and presumably aspherical) explosions of massive stars, known as 
``hypernovae''
are at the origin of the 511 keV emision. This idea has been further elaborated
in Cass\'e et al. (2004, see also contribution by Schanne et al. this volume).
The observed constraints on 511 keV intensity could be satisfied in that case, 
since such explosions are expected to produce large amounts of \Ni \ 
(see Fig. 3,
and SN1998bw in particular) along the rotation axis, thus making easier the
escape of positrons. However, there is no quantitative evaluation yet
of the positron escape fraction in such sources, and any theoretical one
will suffer from the same large uncertainties as the analysis of Chan and 
Lingenfelter (1993) on SNIa; empirical evaluation will require
much larger statistical samples and longer observations of hypernovae
than presently available.
Independently of that, the idea has several 
shortcomings: it is difficult to imagine how the positrons of
a single (or a few) explosion(s) could fill the bulge;  and it is
statistically improbable that such
events are not detected in the Milky Way disk (for instance in the
molecular ring), where the star formation rate is much higher than in
the galactic center or bulge and where all conditions for positron
slow down and annihilation are fulfilled.

Finally, annihilation of a rather special kind of light 
dark matter particles has been recently proposed as the source of
galactic positrons (Boehm et al. 2003, contribution by
Cass\'e et al. this volume). The proposed particles are quite
light (in the 1-100 MeV range) so that their annihilation does not
produce undesirable high energy gamma-rays, and in that respect
they do not correspond to the most commonly discussed dark
matter candidate, which has mass in the GeV to TeV range. Moreover, rather
special properties are required for such light particles to
justify why they have  escaped detection up to now in accelerators
such as the LEP. It is hard to evaluate the plausibility
of that hypothesis, since the required properties of the source
(i.e. density profile, annihilation cross-section) are completely
unknown/unconstrained; in fact, the observed properties of the
511 keV emission (intensity and density profile) are used in
Boehm et al. (2003) in order to derive the properties of the
dark matter source of positrons.

Further observations by INTEGRAL will help, since it is expected that
the Galactic disk will sooner or later 
manifest itself in the light of 511 keV photons. Premiminary hints for
disk 511 keV emission were found in {\it OSSE}'s data (e.g. 
Kinzer et al. 2001, Milne et al. 2002).
The observed disk emission of \Al \ at 1.8 MeV puts a lower limit on the
positron production rate of the disk; observations will establish
whether those positrons annihilate locally (i.e. whether the steady-state
assumption is locally valid) or whether they escape
in the halo to annihilate at higher latitudes.  
In that respect, the Cygnus region,
one of the most prominent ``hot-spots'' in {\it COMPTEL}'s 1.8 MeV map
(see Fig. 7) is a most promising target.

{\it Acknoweledgements:} I am grateful to R. Diel, N. Guessoum, D. Hartmann, 
P. Jean, J. Kn\"odlseder, M. Leising, F. Mirabel, E. van den Heuvel,
P. von Ballmoos, for enlightening discussions during the Workshop. I am
also grateful to the organisers for financial support.

\def\aj{AJ \ }
\def\apj{ApJ \ }
\def\apjs{ApJS \ }
\def\aa{A\&A \ }
\def\aas{A\&AS \ }

\end{document}